\documentclass[prb,showpacs,epsf]{revtex4}
\usepackage{graphicx}
\begin{document}
\addtolength{\baselineskip}{0.8\baselineskip}
\normalsize
\title{Rashba coupling in quantum dots: exact solution}
\author{E. Tsitsishvili$^{(1)}$, 
G.S. Lozano$^{(2,3)}$ and A.O. Gogolin$^{(2)}$}
\affiliation{$~^{(1)}$ Institut f\"{u}r Theorie der Kondensierten 
Materie, Universit\"{a}t Karlsruhe, D-76128 Karlsruhe, Germany;\\
$~^{(2)}$ Department of Mathematics, Imperial College London,
180 Queen's Gate, London SW7 2BZ, United Kingdom;\\
$~^{(3)}$ Departamento de F\'{\i}sica, FCEyN, Universidad de Buenos Aires,
cc 1428, Buenos Aires, Argentina}

\begin{abstract}

We present an analytic solution to the problem
of the Rashba spin-orbit coupling in semiconductor
quantum dots. We calculate 
the exact energy spectrum, wave-functions, and
spin--flip relaxation times.
We discuss various effects inaccessible via perturbation
theory. In particular, we find that the effective gyromagnetic
ratio is strongly suppressed by the spin-orbit coupling.
The spin-flip relaxation rate has a maximum as a function
of the spin-orbit coupling and is therefore suppressed
in both the weak- and strong coupling limits.     

\end{abstract}
\pacs{71.70.Ej, 73.21.La, 72.25.Rb}
\keywords{ spin-orbit coupling, quantum dots, spin relaxation}
\maketitle

In recent years, spin-orbit (SO) effects in quantum
dots attracted much attention, as
it has become clear that these
effects play a crucial role in the
novel field of spintronics \cite{spintronics}.

The combination of a confined geometry and
SO coupling has interesting consequences for
the electron spectrum \cite{spectrum}.
Since future quantum computation devices would
have to control coherent spin states over
sufficiently long time--scales \cite{loss1},
it is important to understand  spin relaxation mechanisms, 
most of which are rooted in the SO coupling.

SO interactions can arise in quantum dots by various
mechanisms related to electron confinement and symmetry 
breaking and are generally known as the Rashba term \cite{rashba} 
and the Dresselhaus term \cite{dresselhaus}.
For most experimental realisations,
quantum dots can be described as effectively two--dimensional 
systems in a confining potential which is usually 
modelled as hard-wall or harmonic confinement. 

To our knowledge all existing theoretical studies of
SO effects in such systems rely on perturbative schemes 
or numerical simulations. 
The purpose of this paper is to 
provide an exact solution 
of the quantum mechanical problem of
the combined effects of the SO coupling,
confinement, and magnetic field.

In the bulk (i.e. without confinement),
the problem was solved in the original
paper \cite{rashba}, see also \cite{br},
while the effects of the confining potential
(but without SO coupling) were studied in Ref.\cite{belgians}. 
We shall show here how these exact solutions
can be combined and generalised. 

The one-particle Hamiltonian describing 
an electron in a two-dimensional quantum dot is of the form:
\begin{equation}
H=\frac{{\bf p}^2}{2m}+V(\rho)+\alpha_R(p_x \sigma_y-p_y\sigma_x) 
+\frac{1}{2}g\mu_B B\sigma_z\;,
\label{ham}
\end{equation}
where $m$ is the effective electron mass, $g$ is the 
effective gyromagnetic ratio, and $\mu_B$ is the Bohr magneton. 
A constant magnetic field $B$
is introduced via the Zeeman term above and the Peierls substitution,
${\bf p}= -i{\bf \nabla} -\frac{e}{c}{\bf A}$ (we use the
axial gauge, $A_\rho=0$ and $A_\varphi=B\rho/2$);
$\alpha_R$ is the strength of the SO coupling, the Pauli
matrices are defined as standard, the electron charge is $e=-|e|$, 
 and we set $\hbar=1$.
In (\ref{ham}), $V(\rho)$ is the (symmetric) confining potential.
In this paper we will consider a hard-wall confining potential, i.e.
$V(\rho)=0$ for $\rho<R$ and $V(\rho)=\infty$ for $\rho>R$,
$R$ being the radius of the dot.
We have  included the Rashba term rather than the Dresselhaus
term, which would be of the form $\alpha_D(p_x\sigma_x-p_y\sigma_y)$. 
(The Rashba term maybe dominant, since the coupling strength $\alpha_R$ 
can be varied by system design, e.g.,  values of  
$\alpha_R \simeq 6 \times 10^{-9}$eV cm were reported for InAlAs/InGaAs 
structures \cite{cui}, whereas the typical values of $\alpha_D$ are about
$10^{-9}$eV cm \cite{nazarov}.)
These terms transform into each other under the spin rotation: 
$\sigma_x\leftrightarrow\sigma_y$, $\sigma_z\leftrightarrow-\sigma_z$.
So our results will only need a trivial modification in the case
when a solo Dresselhaus term is present.

Hamiltonian (\ref{ham}) commutes with the 
$z$--projection of the total momentum operator 
$
j_z=l_z+\frac{1}{2}\sigma_z;$
$l_z=-i\partial_\varphi 
$
(assuming the axial gauge). The operator $j_z$ is therefore
conserved.  
The eigenfunctions of the total momentum operator, with a
half--integer eigenvalue $j$, are of the   form:
\begin{equation}
\psi_j(\rho,\varphi)=(e^{ i (j-1/2) \varphi} f_j(\rho),
e^{i (j+1/2) \varphi} g_j(\rho)).
\end{equation}
In zero field there is an additional
symmetry $j_z\to-j_z$ related to time inversion. The states
with the projections of momenta equal to $j$ and $-j$ are
Kramers doublets. 
Introducing the operators,
$
\nabla^{(B)}_{\pm,j}=\pm\frac{d}{d\rho}-\frac{j}{\rho}-\frac{eB}{c}\rho\;,
$
and 
\[
\Delta^{(B)}_{j}=\frac{1}{\rho}\frac{d}{d\rho}\left(\rho\frac{d}{d\rho}\right)
-\frac{1}{\rho^2}\left(j-\frac{eB}{2c}\rho\right)^2\;,
\]
the Schr\"{o}dinger equation becomes,
\begin{widetext}
\begin{equation}\label{system}
\begin{array}{c}
\Delta^{(B)}_{j-1/2}f_j+2m\left[E-\frac{1}{2}g\mu_B B-V(\rho)\right]f_j
-2m\alpha_R \nabla_{-,j+1/2}^{B}\, g_j=0\;,\\
\Delta^{(B)}_{j+1/2}g_j+2m\left[E+\frac{1}{2}g\mu_B B-V(\rho)\right]g_j
-2m\alpha_R \nabla_{+,j-1/2}^{B}\, f_j=0\;.
\end{array}
\end{equation}
\end{widetext}

We first  assume that the magnetic length
$a_B=\sqrt{c/|e|B}$ is large compared to $R$,
so that the orbital effects
of the magnetic field can be neglected. 
Working  with the dimensionless coordinate $x=\rho/R$,
the system of equations (\ref{system}) becomes 
\begin{equation}\label{systembis}
\begin{array}{c}
(\Delta^{(0)}_{j-1/2}+\epsilon-h)f_j-
\beta_R\nabla^{(0)}_{-,j+1/2}  g_j=0\;,\\
(\Delta^{(0)}_{j+1/2}+\epsilon+h)g_j-
\beta_R\nabla^{(0)}_{+,j-1/2} f_j=0\;,
\end{array}
\end{equation}
supplemented by the boundary conditions $f_j(1)=g_j(1)=0$. We
have introduced two dimensionless parameters, $\beta_R=2\alpha_RmR$
and $h=mg\mu_BR^2B$, characterising the strength of the SO
coupling and the Zeeman term, respectively. The energy 
parameter is $\epsilon=2mER^2$. 

In the absence of the Rashba term and confinement potential,
the solutions regular at the origin are 
$f_j(x)\sim J_{j-1/2} (kx)$ and $g_j(x)\sim J_{j+1/2}(kx)$
with $k^2=\epsilon\pm h$, where $J_l(x)$ are the Bessel functions.
Due to the standard recurrence relations
\begin{equation}
\left(\frac{d}{dx}+\frac{j\pm1/2}{x}\right)J_{j\pm1/2}(kx)=
k J_{j \mp1/2}(kx)\;,
\end{equation}
the Rashba term simply acts as rising or lowering operator on the
Bessel functions basis \cite{rashba}. Therefore the following
ansatz
\begin{equation}\label{ansatz}
(f_j(x),g_j(x))=
(d_1J_{j-1/2}(kx),d_2J_{j+1/2}(kx))
\end{equation}
solves the bulk problem in the presence of the SO
coupling, provided that the
coefficients $d_{1,2}$ satisfy the eigenvalue equation:  
\begin{equation}\label{eigen}
\left[
\begin{array}{cc}
k^2 - \epsilon +h  & -\beta_R k \\
-\beta_R k & k^2 -\epsilon-h
\end{array}
\right]\left[
\begin{array}{c}
d_1\\
d_2
\end{array}
\right]\;
=0.
\end{equation}

When considering the electron confined to the disk, it is seemingly
impossible to impose the vanishing boundary conditions on the
ansatz (\ref{ansatz}) as Bessel functions with different indices
are involved. 
Note, however, that 
as long as either $\beta_R$ or $h$ is non--zero,
the bulk spectrum has two branches:
$
\epsilon=k^2\pm\sqrt{\beta_R^2k^2+h^2}\;.
$
Therefore for a given value of $\epsilon$ there
are, in fact, two non-trivial solutions for the momentum $k$, 
\[
k_{\pm}^2=\frac{ (2\epsilon+\beta_R^2) \pm 
\sqrt{\beta_R^4+4\epsilon\beta_R^2+4h^2}}{2}\;.
\]
We choose the amplitude ratios as
$d_1^+/d_2^+=\alpha_+$ and $d_2^-/d_1^-=\alpha_-$, where
$
\alpha_{\pm}(\epsilon,\beta_R,h)= 
\beta_R k_{\pm}/( k^2_{\pm}-\epsilon \pm h)\;.
$
We are now able to satisfy the boundary conditions by 
combining these two linearly independent degenerate
solutions: 
\begin{eqnarray}
& &(f_j(x),g_j(x))=d^+(\alpha_+ J_{j-1/2}(k_+x),J_{j+1/2}(k_+x)) 
\nonumber
\\
& & +d^- (J_{j-1/2}(k_-x),
\alpha_-J_{j+1/2}(k_-x))\;.
\end{eqnarray}
The boundary conditions lead to the eigenvalue equation 
\[
F J_{j+1/2}(k_+)J_{j-1/2}(k_-)
+J_{j+1/2}(k_-)J_{j-1/2}(k_+)=0\;,
\]
where  the function 
$
F=-\beta_R^2k_+k_-/((k_+^2-\epsilon+h)
(k_-^2-\epsilon-h))\;.
$
Notice  that in the energy region $|\epsilon|<h$,
$k_-$ becomes purely imaginary $k_-\to i\kappa_-$, 
and 
the respective Bessel function
becomes a modified one, $I(\kappa_-x)$.

\begin{figure}
\includegraphics[width=7cm,height=4.5cm]{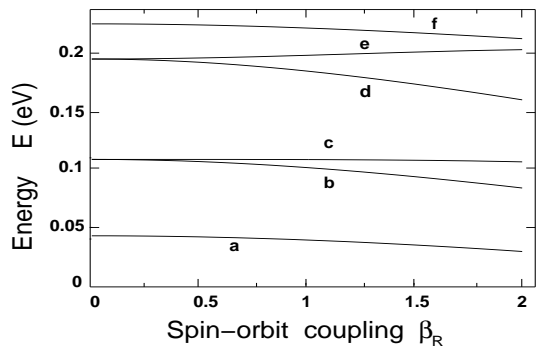}
\caption{Rashba splittings. Energy as a function of $\beta_R$ for the states 
$(1/2,0)$ (a), $(3/2,0)$ (b), $(1/2,1)$ (c), $(5/2,0)$ (d), $(3/2,1)$ (e), 
and $(1/2,2)$ (f).  ($R=10nm$, $B=0$.)}
\label{figp1}
\end{figure}

In the absence of the Zeeman term we have
$k_{\pm}=\sqrt{\epsilon+\beta_R^2/4} \pm\beta_R/2$
and $d_1^{\pm}=\pm d_2^{\pm}$ , so that
$F=1$ in the above equation. In this case,
the equation is invariant under the change 
$j\rightarrow -j$ (Kramers degeneracy).
At $\beta_R=0$ all states with $l\neq 0$ are four-fold
degenerate, while $l=0$ states are doubly degenerate.
According to the standard analysis,
the SO coupling splits all the $l\neq0$ states into
two Kramers doublets with $j=l+1/2$ and $j=l-1/2$,
while $l=0$ states naturally remain Kramers doublets.
The specifics of the Rashba term is that, at small 
$\beta_R$, the SO splitting
starts at the order $\beta_R^2$.
The evolution of the first few energy levels with the
parameter $\beta_R$ is shown in Fig.~\ref{figp1}.
(Values of $m \approx 0.05 m_e$ and 
$g \approx -0.4$, typical for $\rm A_3B_5$ structures, are used.)
We label the energy eigenstates
as $(j,n)$ where $n$ is a non-negative integer such that  
$E_{j,n}<E_{j,n+1}$ at $\beta_R=0$.

\begin{figure}
\begin{center}
\includegraphics[width=7cm,height=6cm]{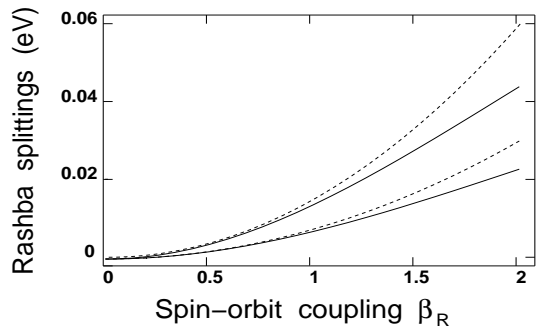}
\vspace{-1.3cm}
\caption{\small Rashba splittings. Energy differences $E(1/2,1)-E(3/2,0)$  and $E(3/2,1)-E(5/2,0)$
as a function of $\beta_R$.  Dashed lines 
correspond to  the same quantities
calculated at the second-order of perturbation theory.
($R=10nm$, $B=0$.) }
\label{figp2}
\end{center}
\end{figure}

A comparison of the exact splittings to the perturbative
ones is shown in Fig.~\ref{figp2}. As one can see from this
figure, the perturbation theory seriously (by 20-30\%) 
overestimates Rashba splittings for realistic values 
of the parameter $\beta_R$.

Upon inclusion of the Zeeman term, all Kramers doublets are
also split so that all the degeneracy is completely lifted.
Because of inherently
small values of the gyromagnetic ratio $g$ in the 
semiconductor quantum dots, for realistic magnetic fields
the Zeeman splittings are small  ($10^{-1}$--$10^{-2}$ meV )
in comparison to the characteristic energy separation between
the levels. Because of the same reason, all the Zeeman 
splittings can be regarded as linear in the magnetic field.
So, for the Zeeman splitting of the $j$'s eigenstate
we may write
$
\delta\epsilon_j=2h f_j(\beta_R)\;,
$
where the function $f_j(\beta_R)$ [$f(0)=1$] 
plays the role of an effective gyromagnetic ratio,
which non-trivially depends on $\beta_R$.
Indeed, expanding the eigenvalue equation in $h$, we find the
following analytic formula for the gyromagnetic factor:
\[
f_j(\beta_R)=t
\left[\frac{J'_{j-1/2}(k_+)}{J_{j-1/2}(k_+)}
-\frac{J'_{j+1/2}(k_+)}{J_{j+1/2}(k_+)}
-(k_+ \leftrightarrow k_-)\right]^{-1},
\]
where the prefactor is $t=4(\epsilon+\beta_R^2/4)/(\beta_R\epsilon)$;
$\epsilon$ and $k_\pm$ are zero--field solutions.
Numerical calculations show that $f(\beta_R)$  is suppressed
by about 50\% when $\beta_R$ reaches 1.5.
Increasing $\beta_R$ suppresses the
Zeeman splitting because  the
SO coupling entangles the spin degree of 
freedom, making it more difficult to polarise.  

\begin{figure}
\begin{center}
\includegraphics[width=7cm,height=4.8cm]{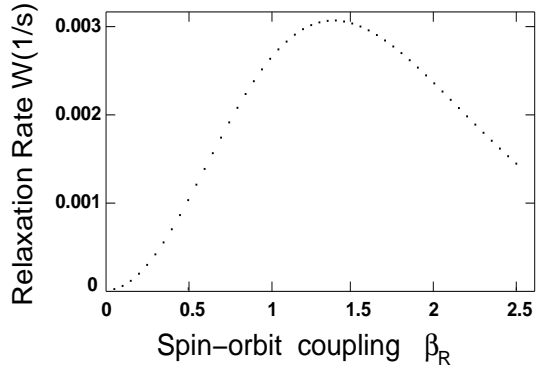}
\caption{\small{The ground-state spin-flip relaxation rate 
$W$ as a function of $\beta_R$ for $B=1.5$T and $R=10$ nm.}}
\label{Wbeta}
\end{center}
\end{figure}

The SO coupling is the main intrinsic mechanism for 
electron spin-flip transitions 
in quantum dots \cite{nazarov}. 
In previous calculations of the spin--flip rates,
the SO coupling was considered as a perturbation, 
so that the electron spin and angular momentum were 
assumed to be independently conserved.
In the full theory this is not the case.
The `spin-flip' transitions in fact occur between
the states $j$ and $-j$. 
No such transition is possible
within a  degenerate Kramers doublet (Van Vleck cancellation).
In the external magnetic field, the states  $j$ and  $-j$ are 
split by the Zeeman interaction. 
The SO coupling allows then for phonon assisted 
transitions between the Zeeman sublevels (of a given Kramers
doublet). 
We concentrate on the most interesting case
and consider the spin--flip transition rate between
the Zeeman sublevels of the ground state ($j=\pm 1/2$). 
 Acoustic phonons dominate these processes at low temperatures.
We consider  piezoelectric interaction between the 
electrons and the acoustic phonons. 
The rate for the one--phonon transition within Zeeman
sublevels is given by the Fermi golden rule.
In analogy Ref.\cite{nazarov}, we obtain (at zero temperature): 
\[
W = \frac{\Bigl(e\;h_{14}\;R\Bigr)^2\;(\Delta E_Z)^3}
{\rho_0  s^5}\;F^2\;K,
\]
\begin{equation}\label{eqF}
F= \int_0^{2 \pi}\;\int_0^1\; \cos \phi \;\psi_{-1/2}^{*}(x, \phi)\;
\psi_{1/2}(x, \phi) x^2 dx \;d\phi
\end{equation}
where $s^{-5} = 3/2 s_l^5 + 1/s_t^5$, $K = 8/105 \pi$, $\Delta E_Z$ is 
the energy difference between 
the states involved, $h_{14}$ is the non-zero component of the piezo-tensor and 
$\rho_0$
is the mass density.
The exact evolution of the spin--flip transition rate with the parameter
$\beta_R$ for fixed $h$ is shown in Fig.~\ref{Wbeta}. 
We use typical material parameters for GaAs-type
structures \cite{rem}. An interesting feature
of the exact solution is the emergence of a maximum
in the transition rate as a function of the spin-orbit
coupling. The physical explanation 
is that while the Zeeman energy splitting
decreases with $\beta_R$, the electron-phonon matrix
element saturates.
For  small $\beta_R$ and $h$, 
we obtain $F\sim \beta_R h$  so that the transition rate is 
$W \sim h^5 \beta_R^2$, in accordance
with the perturbative result of Ref.\cite{nazarov}.

So far we have assumed that $a_B>R$, 
which is the case for most experimental
set-ups. 
For bigger dots and stronger magnetic fields, however, the orbital effects of
the magnetic field can
not be neglected. 
Fortunately, because of the very
nature of the Peierls substitution, which has to be performed
both in the kinetic energy term and in the SO
term, the above analytic solution can be 
generalised to this case. 

Using the dimensionless variable
$\xi=\rho^2/(2 a_B^2)$ we propose the ansatz
\[
(f_j(\xi),g_j(\xi))=(d_1 \Phi(e_0,j,\xi),d_2 \Phi(e_0+1,j+1,\xi))\;,
\]
where
\begin{eqnarray}
\Phi(e_0,j,\xi)&=&c_j \xi^{\frac{|l|}{2}} M (-e_0+\frac{l+|l|}{2},
|l|+1,\xi) \nonumber \\
c_j &=&\left( 
\frac{\Gamma(e_0-\frac{l-|l|}{2}+1)}
{\Gamma(e_0-\frac{l+|l|}{2}+1)} \right)^{\frac{1}{2}} \frac{(-1)^{\frac{-l+|l|}{2}}}{\Gamma(|l|+1)}
\nonumber
\end{eqnarray}
with $l=j-\frac{1}{2}$
and  $M(a,c;\xi)$ is the confluent hypergeometric 
function (see \cite{bateman}). 
\begin{figure}
\begin{center}
\includegraphics[width=8cm,height=5.2cm]{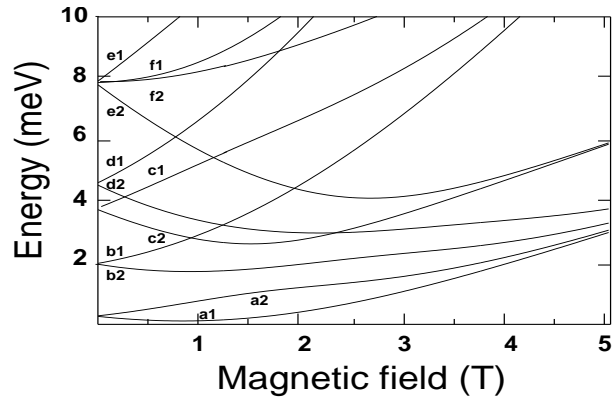}
\caption{\small The spectrum in the orbital field. 
Energy as a function of magnetic field for the states 
$(1/2,0)$ (a1), $(3/2,0)$ (b1), $(1/2,1)$ (c1), $(5/2,0)$ (d1), 
$(3/2,1)$ (e1), 
and $(1/2,2)$ (f1), $(-1/2,0)$ (a2), $(-3/2,0)$ (b2), $(-1/2,1)$
(c2), $(-5/2,0)$ (d2), $(-3/2,1)$ (e2) and $(-1/2,2)$ (f2). 
($R=50nm$, $\beta_R=3.35$.) }
\label{prueba}
\end{center}
\end{figure}
The factors $c_j$ are inspired by the normalisation 
in the standard Landau problem  and $e_0$ is
the `energy' parameter to be determined.
As before, we find two non-trivial solutions for it:
\[
e_0^{\pm}=\frac{2e-1+\gamma^2 \pm \sqrt{ (\gamma^2+1)^2+4 
e\gamma^2+4(s^2-s)}}{2}\;.
\]
We have introduced dimensionless parameters:
$e=E/\omega_c-1/2$ ($\omega_c=|e|B/mc$) 
for the energy (not to be confused
with the electron charge), 
$\gamma=\alpha_d (2m/\omega_c)^{1/2}$
for the SO coupling, and $s=gm/(4m_e)$ for the
Zeeman coupling. Here $m_e$ is the electron mass. 
The boundary conditions then lead to the eigenvalue equation:
\[
\frac{ 
\Phi(e_0^+,l,\xi_0)
\Phi(e_0^-+1,l+1,\xi_0)}
{\Phi(e_0^-,l,\xi_0)
\Phi(e_0^++1,l+1,\xi_0)}
=
\frac{(e_0^{-}+1)^\frac{1}{2}}
{(e_0^{+}+1)^\frac{1}{2}}
\frac{
e_0^+-e+s}
{
e_0^--e+s},
\]
where $\xi_0=R^2/(2a_R^2)$.
This equation provides all the information about the 
energy spectrum of the system.
We have
investigated it numerically. The results are shown in Fig.~\ref{prueba}. 
Note that the Zeeman splittings are still small for
realistic fields. The Kramers doublets therefore survive
the orbital field as long as there is no SO
coupling. It is the combined effect of the orbital
field and the Rashba term that lifts the Kramers degeneracy.


To conclude, we presented an analytic solution to the problem of an
electron in a quantum dot in the presence of both the 
magnetic field and SO coupling. 
We calculated 
various quantities of physical interest. 
For realistic parameters, 
the Rashba energy splittings are overestimated
in perturbation theory. There is also a strong suppression of the
effective gyromagnetic ratio by the SO coupling.
The spin-flip relaxation rate has a maximum as a function of
the SO coupling, a prediction that would be interesting to
verify experimentally.  Inclusion of 
the orbital magnetic field gives rise to a rich
magneto-optical spectrum. We hope that our solution can
be used in future research for obtaining further interesting
results on the SO effects in quantum dots.


We are grateful to Levitov, who has independently
arrived at a similar solution with Rashba \cite{levitov}, 
for interesting discussions.
G.L.'s and A.O.G.'s research is supported by the EPSRC
grants GR/N19359 and GR/R70309 and the EU training network
DIENOW.
E.T.'s  research is supported by the Centre for Functional Nanostructures
 of the Deutsche Forschungsgemeinschaft  within project A2.

{\noindent \em Note added} After completion of this work we learnt that the 
particular case of zero magnetic field was analised by Bulgakov and
Sadreev~\cite{sad}.


\begin{thebibliography}{99}
\bibitem{rem} Typical material parameters are: 
$e h_{14} \approx 1.2\times 10^7$eV/cm,
$s_{t,l} \approx 3-5\times 10^5$cm/sec, $\rho_0 \approx 5$g/cm$^3$.
\end{thebibliography}
\end{document}